\documentclass[final,times,twocolumn]{elsarticle}
\usepackage{graphicx,amssymb,amsmath,multirow,dcolumn,bm,latexsym,soul}
\newcolumntype{K}[1]{>{\centering\arraybackslash}p{#1}}

\journal{Physics Letters B}
\begin{document}
\newcommand{\be}{\begin{equation}}
\newcommand{\ee}{\end{equation}}
\newcommand{\bq}{\begin{eqnarray}}
\newcommand{\eq}{\end{eqnarray}}

\begin{frontmatter}
\title{Cosmological impact of microwave background temperature measurements}
\author[inst1]{L. Gelo}\ead{lea.gelo@student.ung.si}
\author[inst2,inst3]{C. J. A. P. Martins\corref{cor1}}\ead{Carlos.Martins@astro.up.pt}
\author[inst4]{N. Quevedo\fnref{present1}}\ead{n.quevedo.cabrera@student.rug.nl}
\author[inst2,inst5]{A. M. M. Vieira\fnref{present2}}\ead{ana.mafalda.vieira@sapo.pt}
\address[inst1]{School of Science, University of Nova Gorica, Vipavska cesta 11c, Ajdov\v{s}\v{c}ina, Slovenia}
\address[inst2]{Centro de Astrof\'{\i}sica da Universidade do Porto, Rua das Estrelas, 4150-762 Porto, Portugal}
\address[inst3]{Instituto de Astrof\'{\i}sica e Ci\^encias do Espa\c co, CAUP, Rua das Estrelas, 4150-762 Porto, Portugal}
\address[inst4]{British School of Tenerife, Camino Montijo, 16, 38410 Los Realejos, Santa Cruz de Tenerife, Spain}
\address[inst5]{Col\'egio da Rainha Santa Isabel, Rua do Brasil, 3030-175 Coimbra, Portugal}
\cortext[cor1]{Corresponding author}
\fntext[present1]{Present address: Faculty of Science and Engineering, University of Groningen, PO Box 72, 9700 AB Groningen, The Netherlands}
\fntext[present2]{Present address: Faculdade de Ci\^encias da Universidade de Lisboa
Campo Grande, 1749-016 Lisboa, Portugal}

\begin{abstract}
The cosmic microwave background temperature is a cornerstone astrophysical observable. Its present value is tightly constrained, but its redshift dependence, which can now be determined until redshift $z\sim6.34$, is also an important probe of fundamental cosmology. We show that its constraining power is now comparable to that of other background cosmology probes, including Type Ia supernovae and Hubble parameter measurements. We illustrate this with three models, each based on a different conceptual paradigm, which aim to explain the recent acceleration of the universe. We find that for parametric extension of $\Lambda$CDM the combination of temperature and cosmological data significantly improves constraints on the model parameters, while for alternative models without  a $\Lambda$CDM limit this data combination rules them out.
\end{abstract}

\end{frontmatter}

\section{Introduction}\label{into}

The present-day value of the cosmic microwave background (CMB) temperature has been accurately measured by COBE-FIRAS, $T_0= 2.7255\pm 0.0006$ K \cite{Fixsen}. If the expansion of the Universe is adiabatic, photon number is conserved, and the CMB spectrum was originally a black-body, the CMB temperature will evolve as $T(z)=T_0(1+z)$. This is a robust prediction of standard cosmology, but it is violated in many contexts, from exotic astrophysical processes to particle physics and cosmological models beyond $\Lambda$CDM \cite{Avgoustidis,Euclid}. Deviations from this behaviour would imply new physics, motivating efforts towards measurements at non-zero redshifts.

There are two paths for such measurements: the thermal Sunyaev-Zel'dovich effect at low redshifts, and spectroscopy of suitable molecular or atomic species at higher redshifts. The first measurement (as opposed to an upper limit) was achieved in 2000 \cite{Srianand}, and their number has been steadily increasing. Recently, the redshift range of these measurements has been extended to $z\sim6.34$ \cite{Riechers}.

These measurements have been used to constrain particular models, but there is no detailed study of their constraining power, as compared to that of other background cosmology data sets at similar redshifts. We present the first such study, showing that CMB temperature measurements are as constraining as Type Ia supernova and Hubble parameter data, and have important synergies with them. We discuss three different cosmological models, each with a different motivation.

\section{Background cosmology data}\label{data}

CMB temperature measurements are effectively background cosmology data. Therefore, our benchmark for their constraining power is two other canonical background cosmology data sets. The first is the Pantheon compilation of Scolnic \textit{et al.} \cite{Scolnic,Riess}. This is a 1048 supernova data set, compressed into 6 correlated measurements of $E^{-1}(z)$ (where $E(z)=H(z)/H_0$ is the dimensionless Hubble parameter) in the redshift range $0.07<z<1.5$. The second is the compilation of 38 Hubble parameter measurements by Farooq \textit{et al.} \cite{Farooq}: this includes both data from cosmic chronometers and from baryon acoustic oscillations. Together, the two sets contain measurements up to redshift $z\sim2.36$, and when using the two in combination we will refer tho this as the cosmological data. The Hubble constant is always analytically marginalized \cite{Anagnostopoulos}.

The CMB temperature constraints are listed in Table \ref{tab1}. At low redshifts they come from the thermal Sunyaev-Zel'dovich (SZ) effect, specifically from 815 Planck clusters in 18 redshift bins \cite{Hurier} and 158 SPT clusters in 12 redshift bins \cite{Saro}. At high redshifts we have spectroscopic measurements at various wavelengths and using several molecular or atomic species. The recent work of \cite{Klimenko} updates some earlier analyses \cite{Srianand0,Noterdaeme1,Noterdaeme2,Krogager}, specifically including the contribution of collisional excitation in the diffuse interstellar medium to the excitation temperature of the tracer species. The highest redshift measurement is at $z\sim6.34$, but must of them are until $z\sim2.5$, so the redshift range of the cosmological and CMB data sets is comparable. In our analysis we do not need the present-day value of the CMB temperature: $T_0$ is analytically marginalized, exactly as is done for $H_0$.

We report constraints for three classes of models, from on a standard statistical likelihood analysis, based on Matlab and Python codes that have been custom-built for this work, but validated against each other and also by comparison with analogous results for the same models in the literature, when these exist.

\begin{table}
\begin{center}
\caption{Measurements of the CMB temperature, with one-sigma uncertainties. The top and bottom parts of the table correspond to SZ and spectroscopic measurements respectively.}
\label{tab1}
\begin{tabular}{| c | c | c |}
\hline
Redshift & $T(z)$ & Reference \\
\hline
$0.037$ & $2.888\pm0.041$ & \cite{Hurier} \\
$0.072$ & $2.931\pm0.020$ & \cite{Hurier} \\
$0.125$ & $3.059\pm0.034$ & \cite{Hurier} \\
$0.129$ & $3.01_{-0.11}^{+0.14}$ & \cite{Saro} \\
$0.171$ & $3.197\pm0.032$ & \cite{Hurier} \\
$0.220$ & $3.288\pm0.035$ & \cite{Hurier} \\
$0.265$ & $3.44_{-0.13}^{+0.16}$ & \cite{Saro} \\
$0.273$ & $3.416\pm0.040$ & \cite{Hurier} \\
$0.332$ & $3.562\pm0.052$ & \cite{Hurier} \\
$0.371$ & $3.53_{-0.14}^{+0.18}$ & \cite{Saro} \\
$0.377$ & $3.717\pm0.065$ & \cite{Hurier} \\
$0.416$ & $3.82_{-0.15}^{+0.19}$ & \cite{Saro} \\
$0.428$ & $3.971\pm0.073$ & \cite{Hurier} \\
$0.447$ & $4.09_{-0.19}^{+0.25}$ & \cite{Saro} \\
$0.471$ & $3.943\pm0.113$ & \cite{Hurier} \\
$0.499$ & $4.16_{-0.20}^{+0.27}$ & \cite{Saro} \\
$0.525$ & $4.380\pm0.120$ & \cite{Hurier} \\
$0.565$ & $4.075\pm0.157$ & \cite{Hurier} \\
$0.590$ & $4.62_{-0.26}^{+0.36}$ & \cite{Saro} \\
$0.619$ & $4.404\pm0.195$ & \cite{Hurier} \\
$0.628$ & $4.45_{-0.23}^{+0.31}$ & \cite{Saro} \\
$0.676$ & $4.779\pm0.279$ & \cite{Hurier} \\
$0.681$ & $4.72_{-0.27}^{+0.39}$ & \cite{Saro} \\
$0.718$ & $4.933\pm0.371$ & \cite{Hurier} \\
$0.742$ & $5.01_{-0.33}^{+0.49}$ & \cite{Saro} \\
$0.783$ & $4.515\pm0.621$ & \cite{Hurier} \\
$0.870$ & $5.356\pm0.617$ & \cite{Hurier} \\
$0.887$ & $4.97_{-0.19}^{+0.24}$ & \cite{Saro} \\
$0.972$ & $5.813\pm1.025$ & \cite{Hurier} \\
$1.022$ & $5.37_{-0.18}^{+0.22}$ & \cite{Saro} \\
\hline
$0.89$ & $5.08\pm0.10$ & \cite{Muller} \\
$1.73$ & $7.9^{+1.7}_{-1.4}$ & \cite{Klimenko} \\ 
$1.77$ & $6.6^{+1.2}_{-1.1}$ & \cite{Klimenko} \\ 
$1.78$ & $7.2\pm0.8$ & \cite{Cui} \\
$1.97$ & $7.9\pm1.0$ & \cite{Ge} \\ 
$2.04$ & $8.6^{+1.9}_{-1.4}$ & \cite{Klimenko}  \\ 
$2.34$ & $10\pm4$ & \cite{Srianand} \\
$2.42$ & $9.0^{+0.9}_{-0.7}$ & \cite{Klimenko} \\ 
$2.53$ & $9.8^{+0.7}_{-0.6}$ & \cite{Klimenko} \\ 
$2.63$ & $10.8^{+1.4}_{-3.3}$ & \cite{Klimenko} \\ 
$2.69$ & $10.4^{+0.8}_{-0.7}$ & \cite{Klimenko} \\ 
$3.02$ & $12.1^{+1.7}_{-3.2}$ & \cite{Molaro} \\ 
$3.09$ & $12.9^{+3.3}_{-4.5}$ & \cite{Klimenko} \\ 
$3.29$ & $15.2^{+1.0}_{-4.2}$ & \cite{Klimenko} \\  
$6.34$ & $23.1^{+7.1}_{-6.7}$ & \cite{Riechers} \\
\hline
\end{tabular}
\end{center}
\end{table}

\section{The Jetzer {\it et al.} model}\label{model1}

This model \cite{Jetzer1,Jetzer2} is based on an earlier idea by Lima \cite{Lima}. It is a decaying dark energy model where photon creation is possible, leading (among others) to a non-standard temperature redshift relation.  We assume flat universes. For sufficiently low redshfts for the radiation to be negligible, the Friedmann equation can be written
\be
E^2(z)=\frac{3\Omega_m-m}{3-m}(1+z)^3+\frac{3(1-\Omega_m)}{3-m}(1+z)^m\,.
\ee
The dark energy decay is parametrized by $m$ (with $m=0$ recovering flat $\Lambda$CDM) and it can be thought of as an effective dark energy equation of state $w_{eff}=m/3-1$.

\begin{figure}
\begin{center}
\includegraphics[width=\columnwidth]{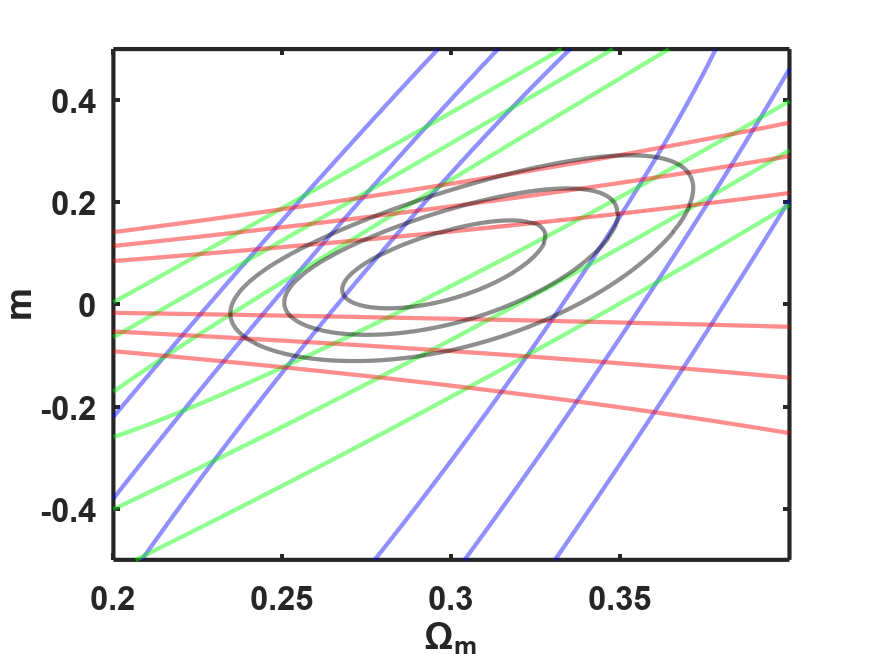}
\includegraphics[width=\columnwidth]{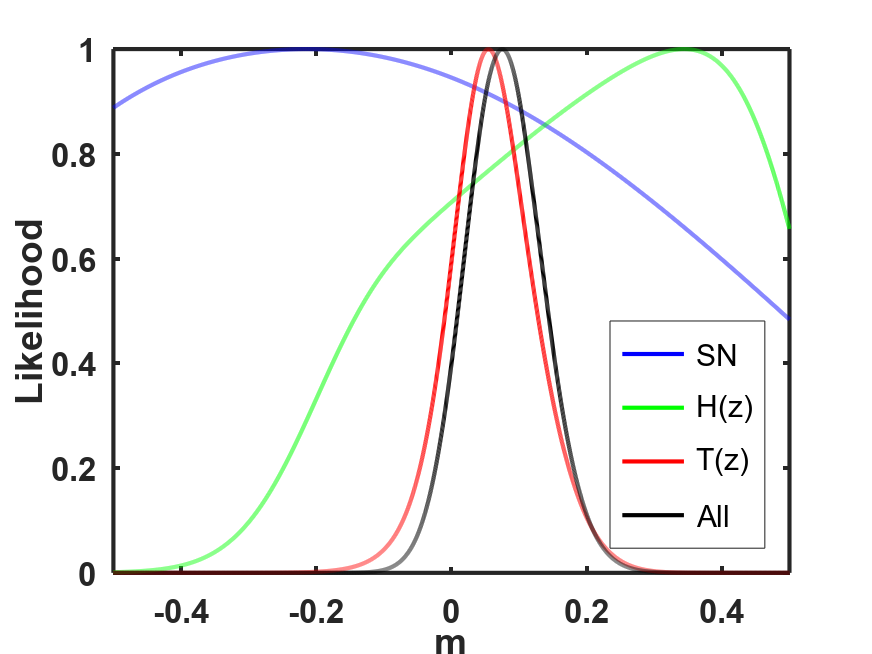}
\includegraphics[width=\columnwidth]{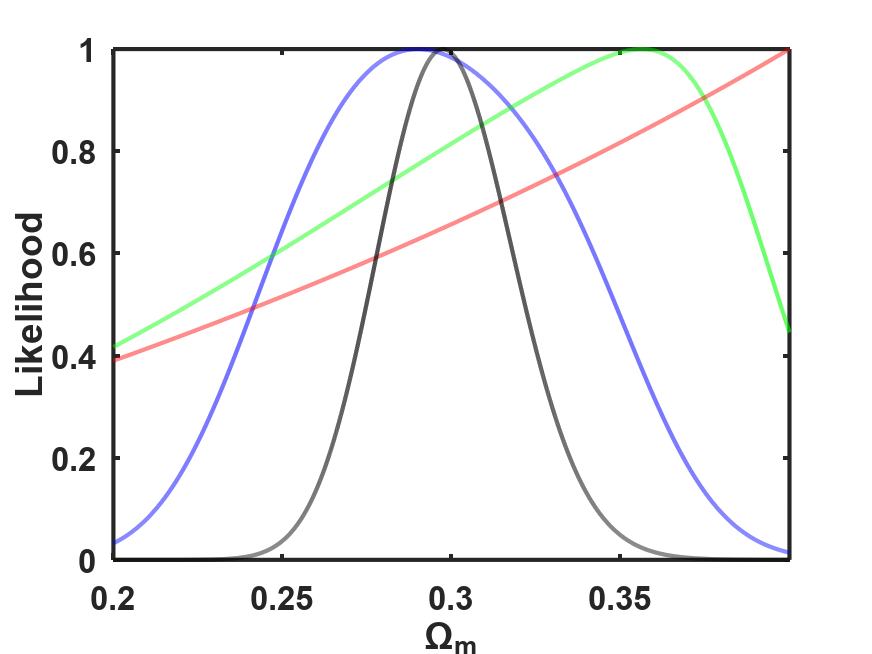}
\end{center}
\caption{\label{fig1}Constraints on the Jetzer \textit{et al.} model with $w_r=1/3$. Blue, green, and red curves show constraints from supernova, Hubble parameter and temperature data, while black ones depict the combined constraints. The top panel shows the one, two and three sigma constraints on the $m$--$\Omega_m$ plane, while the others show the posterior marginalized likelihoods for the individual parameters.}
\end{figure}

The model's coupling to radiation must necessarily be weak, as otherwise it would be already ruled out \cite{Chluba}: such a model can only be a small perturbation on the standard $\Lambda$CDM behaviour. Under these assumptions the temperature-redshift relation can be written
\be
\frac{T(z)}{T_0}=(1+z)^{3w_r}\left[\frac{(m-3\Omega_m)+m(\Omega_m-1)(1+z)^{m-3}}{(m-3)\Omega_m}\right]^{w_r}\,,
\ee
where we also relax the assumption of a strict radiation fluid, allowing it to have a generic constant equation of state $w_r$, with the standard case being $w_r=1/3$.

An early analysis \cite{Jetzer1,Jetzer2}, which using coeval data analogous to the present one (but kept $T_0$ fixed) finds constraints on $m$ of the order of $\sigma_m\sim0.1$ for the standard value of $w_r$, while if $w_r$ is free to vary the constraint on $m$ is of order $\sigma_m\sim0.2$, with $\sigma_{wr}\sim0.03$. On the other hand, \cite{Noterdaeme2}, with additional temperature measurements, obtained $\sigma_m\sim0.075$. Later \cite{Hurier,Saro} both independently obtained $\sigma_m\sim0.05$, for the standard value of $w_r$. A recent analysis \cite{Lucca} of a broader class of models uses the full the Planck data (but not the Table \ref{tab1} data) and finds, for the specific model we study, $\sigma_m\sim0.07$. On the other hand, \cite{Riechers} obtains uncertainties of$\sigma_m\sim0.05$, with $\sigma_{wr}\sim0.01$, but assumes a matter density fixed to the Planck value.

We restrict ourselves to the data introduced in the previous section. Figure \ref{fig1} shows our constraints, for the standard equation of state for radiation, $w_r=1/3$. We separately show the constraints from supernovas, Hubble parameter and temperature data, to highlight how their combination breaks degeneracies. We see that the three data sets are mutually consistent, and the key role of the temperature data in constraining $m$. Using the cosmology data alone one finds $\Omega_m=0.32\pm0.04$ and a weak constraint $m=0.23\pm0.18$, while if using the temperature data alone the matter density is unconstrained while $m=0.06\pm0.05$. For the combined data sets, we find
\bq
\Omega_m&=&0.30\pm0.02\\
m&=&0.07\pm0.05\,,
\eq
which compares favourably with previous constraints---recall that we analytically marginalize $T_0$ as well as $H_0$. The combined data set improves the constraint on the matter density by a factor of two.

The above does assume the standard equation of state for radiation, $w_r=1/3$, but this can be relaxed, in which case the one-sigma constraints from the full data are
\bq
\Omega_m&=&0.32\pm0.03\\
m&=&0.23\pm0.13\\
w_r&=&0.35\pm0.02\,;
\eq
here there are additional degeneracies with weaken the constraint on $m$, while $w_r$ is tightly constrained. Overall, there are no statistically significant deviations from the $\Lambda$CDM behaviour.

\section{The Canuto {\it et al.} model}\label{model2}

The motivation for this model \cite{Canuto1,Canuto2} is the fact that, although the effects of scale invariance vanish in the presence of particles with non-zero rest masses, one may assume that on cosmological scales empty space should still be scale invariant. This assumption ultimately leads to a bimetric theory, with a time-dependent function $\lambda$ playing the role of a scale transformation factor relating the ordinary matter frame to a scale invariant frame. The first of these can be thought of as the atomic (or physical) frame, while the second is a gravitational frame, in which the ordinary Einstein equations would still hold \cite{Canuto2}.

We assume homogeneous and isotropic universes, and a vanishing curvature parameter---previous work shows that relaxing this assumption has no significant impact \cite{Covariant}. We assume a generic power-law function $ \lambda(t)=(t/t_0)^p=x^p$, where $t_0$ is the present age of the universe and we have defined a dimensionless age of the universe, $x$; note that $\lambda(t_0)=1$, so $\Lambda$CDM is recovered for $p=0$. With these assumptions the Friedmann equation is
\be
\left(E(z,x)+\frac{p}{2x}\Omega_\lambda\right)^2=\Omega_m(1+z)^{3(1+w)}x^{-p(1+3w)}+\Omega_{\Lambda}x^{2p}\,,
\label{scalef2}
\ee
where $\Omega_\lambda=2/t_0H_0$; note that there is a consistency condition $(1+p\Omega_\lambda/2)^2=\Omega_m+\Omega_{\Lambda}$.

For numerical convenience this can be re-written as
\bq
E(z,x)&=&\frac{\Omega_\lambda}{2x}\left[-p+\sqrt{N(z,x)}\right]\\
\frac{\Omega_\lambda^2}{4}N(z,x)&=&\Omega_m(1+z)^{3(1+w)}x^{2-p(1+3w)}+\Omega_\Lambda x^{2(1+p)}\,,
\eq
together with
\be
\frac{dx}{dz}=- \frac{x}{1+z}\times\frac{1}{\sqrt{N(z,x)}-p}\,,
\ee
with the initial condition $x=1$ at $z=0$. Finally, the temperature-redshift relation is \be
T(z)=T_0(1+z)x^{-p/2}\,.
\ee

\begin{figure*}
\begin{center}
\includegraphics[width=\columnwidth]{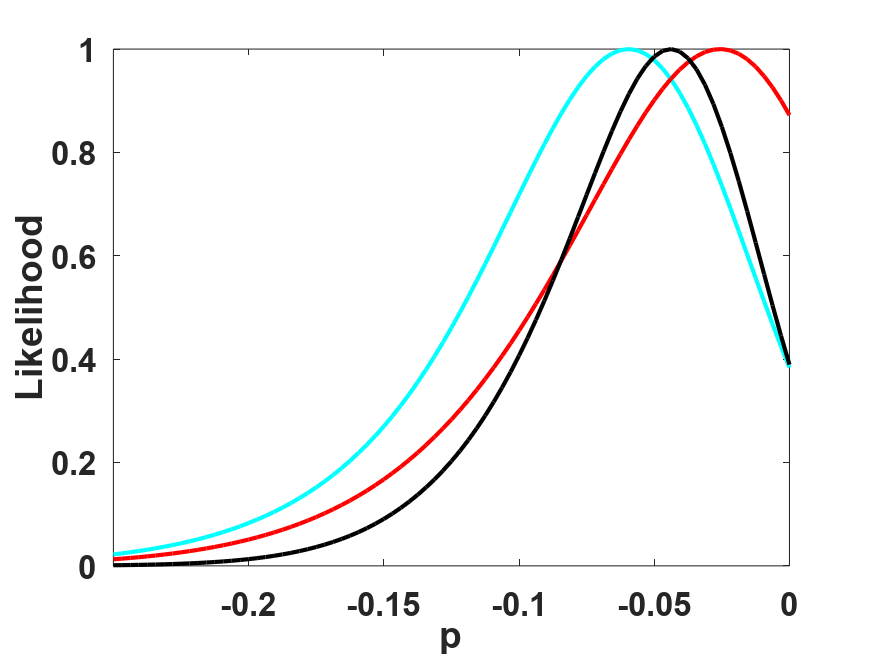}
\includegraphics[width=\columnwidth]{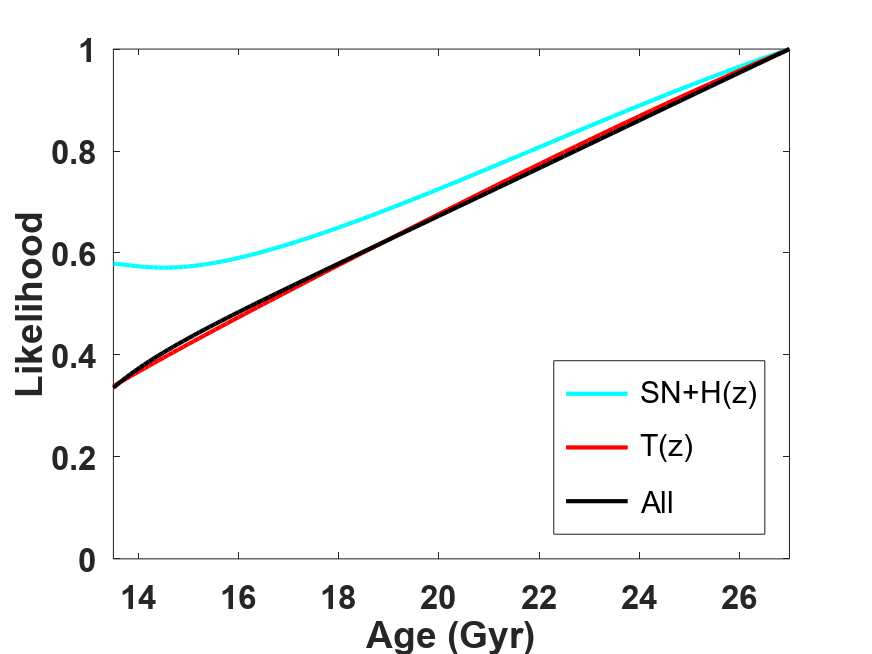}
\end{center}
\caption{\label{fig2}Constraints on the Canuto \textit{et al.} model discussed in the text. Blue, red and black curves show constraints from cosmology, temperature data, and the combined constraints, respectively. The panels show the posterior marginalized likelihoods for $p$ and the matter density.}
\end{figure*}

The model can be studied in two separate contexts. The first assumes $\Omega_\Lambda=0$, making it an alternative to $\Lambda$CDM. However, previous work \cite{Covariant} shows that such a scenario is ruled out: for low-redshift cosmological data one could obtain a statistically reasonable fit for $\Omega_m=0.11\pm0.03$ and a matter equation of state $w_m=0.51\pm0.16$, but such exotic values would be in conflict with higher-redshift data \cite{Planck}. Therefore, in what follows we focus on the second context, where $\Omega_\Lambda\neq0$ and the model is a parametric extension of $\Lambda$CDM.

Figure \ref{fig2} shows our results. We have chosen a uniform prior on $\Omega_\lambda$ corresponding to a present age of the universe from 13.5 to 27 Gyr; the lower limit corresponds to the age of the oldest identified galaxy, GN-z11 \cite{Eleven}, further corroborated by estimates from galaxy clusters \cite{Valcin}. We also assume a standard equation of state for matter $w=0$, and restrict our analysis to $p\le0$, in which case we effectively have a decaying cosmological constant---except if $p=0$, which recovers $\Lambda$CDM. These choices do not significantly impact our results.

From the cosmology data we find the one-sigma constraint $\Omega_m=0.268\pm0.020$ and the two-sigma lower limit $p>-0.18$; the temperature data does not provide a significant constraint on $\Omega_m$ but yields the lower limit $p>-0.16$. For the full data set we find
\bq
\Omega_m&=&0.272\pm0.018\\
p&>&-0.14\,.
\eq
The addition of the temperature data improves the matter density contraint by $10\%$ and the limit on the parameter $p$ by $24\%$. We also note that in this model there is a mild preference for an age of the universe that is larger than the canonical one, as can be seen in the right panel of Fig. \ref{fig2}.

\section{The fractional cosmology model}\label{model3}

Finally, we consider the recently proposed fractional cosmology model \cite{Fractional}. This is based on mathematical formalism of fractional calculus \cite{Tarasov}; for our purposes, it suffices to note that it is a parametric extension of the usual concepts of differentiation and integration.

The model is envisaged as an alternative to $\Lambda$CDM, in the sense that it is assumed to contain no cosmological constant. Nevertheless, we will in what follows include a cosmological constant in the relevant equations. The Friedmann equation now has the form
\be
E^2+\frac{1-\alpha}{H_0t_0} \frac{E}{x}=\Omega_m(1+z)^{3}x^{\alpha-1}+\Omega_\Lambda\,,
\label{fraceq1}
\ee
where $t_0$ and $x$ have the same definitions as the previous section, and $\alpha$ is the fractional calculus parameter: in standard calculus one has $\alpha=1$. Since $E(z=0)=1$, we have $1/(H_0t_0)=(\Omega_m+\Omega_\Lambda-1)/(1-\alpha)$, so can also write
\be
E^2+(\Omega_m-1)\frac{E}{x}=\Omega_m(1+z)^3x^{\alpha-1}+\Omega_\Lambda\,. \label{fraceq2}
\ee
From this we see that for $\Omega_m+\Omega_\Lambda<1$ one needs $\alpha>1$; conversely, $\alpha<1$ would require $\Omega_m+\Omega_\Lambda>1$, otherwise one would have a negative age of the universe.

As in the previous section, we can also write this as
\be
E(z,x)=\frac{1-\Omega_m}{2x}+ \sqrt{\frac{(1-\Omega_m)^2}{4x^2}+\Omega_m(1+z)^3x^{\alpha-1}+\Omega_\Lambda}\,,
\ee
together with
\be
\frac{dx}{dz}=\frac{1-\Omega_m-\Omega_\Lambda}{1-\alpha}\times\frac{1}{(1+z)E(z,x)}\,,
\ee
Finally the temperature redshift relation is
\begin{equation}
    T(z)=T_0(1+z)x^{(\alpha-1)/3}\,;
\end{equation}
these recover the standard $\Lambda$CDM behaviour for $\alpha=1$.

\begin{figure*}
\begin{center}
\includegraphics[width=\columnwidth]{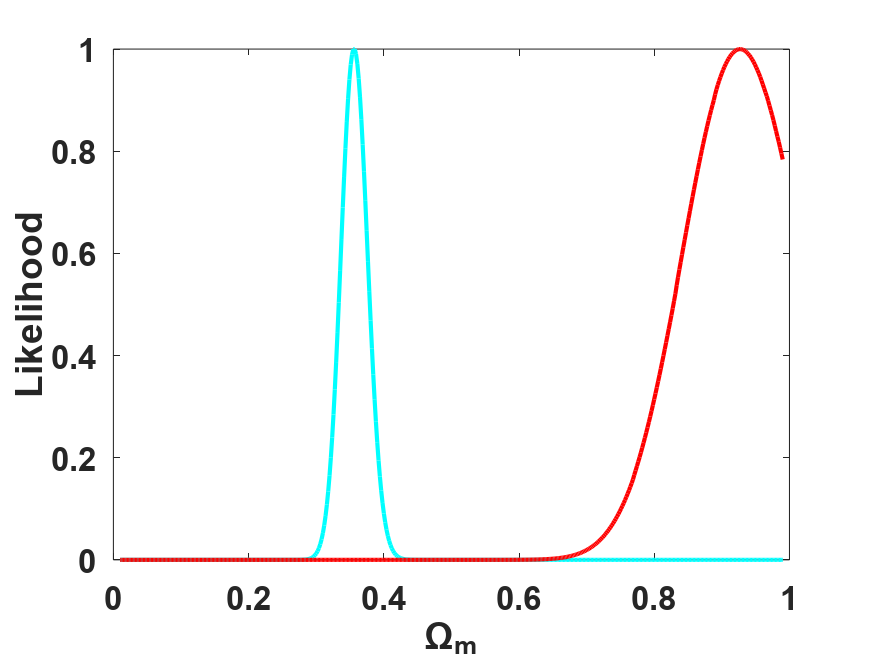}
\includegraphics[width=\columnwidth]{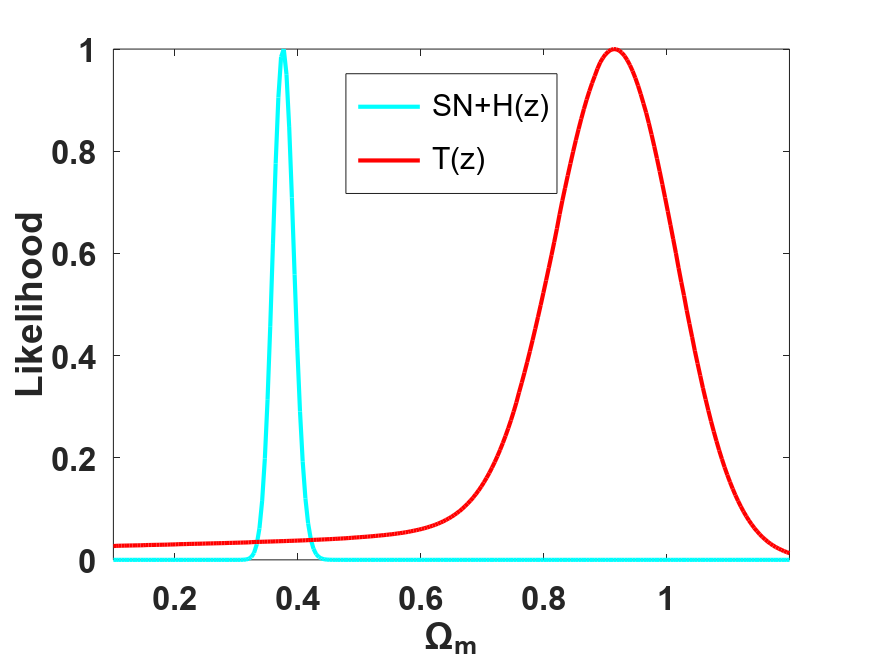}
\end{center}
\caption{\label{fig3}Constraints on the fractional model discussed in the text. Blue and red curves show constraints from cosmology and temperature data, respectively. The panels show the posterior marginalized likelihoods for the matter density; the left panel shows the physical region $\alpha>1$, while the right panel shows the $\alpha<1$ region, which is unphysical (implying a negative age of the universe) unless $\Omega_m>1$. Notice that no model parameters simultaneously provide a good fit to the two data sets}
\end{figure*}

A recent analysis \cite{Fractional}, using the Pantheon supernova and cosmic chronometers data and assuming $\Omega_\Lambda=0$, finds $\alpha=2.8\pm0.2$ and $\Omega_m=0.23\pm0.04$ (implying an age of the universe of $t_0\sim33.6$ Gyr). That work does not use baryon acoustic oscillation data, and treats the Hubble constant as an additional parameter, In our analysis, we do include baryon acoustic oscillation data (which is part of the Hubble parameter compilation), as well as CMB temperature data, and analytically marginalize $H_0$ and $T_0$.

Our results, for $\Omega_\Lambda=0$, are in the left panel of Figure \ref{fig3}. We assume generous uniforms prior for the matter density, $\Omega_m\in]0,1[$, and the fractional calculus parameter, $\alpha\in]1,4]$. The cosmological data leads to $\Omega_m=0.36\pm0.02$ and a fractional calculus parameter as large as possible, but this is incompatible with the temperature data, which leads to the two-sigma lower limit $\Omega_m>0.76$. A large matter density is preferred because it ensures $x\sim1$, approximately preserving the temperature-redshift relation and allowing it to fit the existing data. The model is therefore ruled out. Since the best-fit parameters of the two data sets are mutually incompatible, it makes no statistical sense to combine them in a join likelihood.

The mismatch in best-fit values from cosmology and temperature data also applies to the $\alpha<1$ region. This is shown in the right panel of Figure \ref{fig3}, where we also allowed for $\Omega_m>1$ to enable positive ages of the universe. Nevertheless. constraints on the matter density are almost unchanged---$\Omega_m=0.38\pm0.02$ from cosmology data and $\Omega_m=0.9\pm0.1$ from temperature data---and even in this wider parameter space, the model would be ruled out.

Finally, consider the case with $\Omega_\Lambda\neq0$, which would make the model a parametric extension of $\Lambda$CDM. For simplicity we fix the age of the universe to the standard value \cite{Planck}. The Friedmann equation can then be written
\be
E(z,x)=\frac{1-\alpha}{2x}+ \sqrt{\frac{(1-\alpha)^2}{4x^2}+\Omega_m(1+z)^3x^{\alpha-1}+\alpha-\Omega_m}\,,
\ee
with
\be
\frac{dx}{dz}=-\,\frac{1}{(1+z)E(z,x)}\,,
\ee
and the same temperature-redshift relation. Evidently, $\alpha=1$ corresponds to the $\Lambda$CDM model. The two data sets are now mutually compatible, and the addition of the temperature data improves constraints on the matter density from $\Omega_m=0.39_{-0.08}^{+0.06}$ to $\Omega_m=0.32\pm0.04$, again a significant improvement, while the overall constraint on the fractional calculus parameter is $\alpha=1.048\pm0.043$.

\begin{figure*}
\begin{center}
\includegraphics[width=\columnwidth]{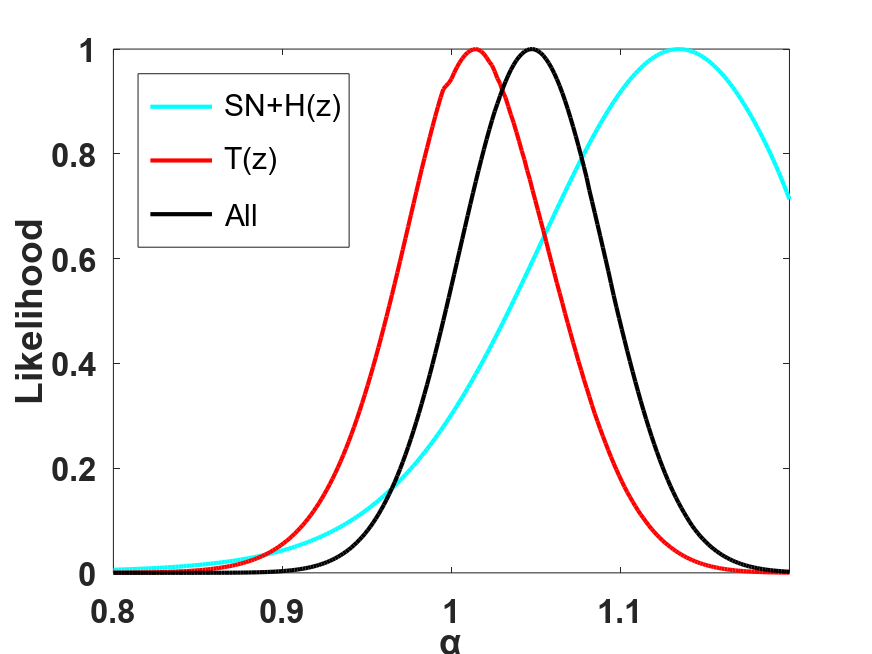}
\includegraphics[width=\columnwidth]{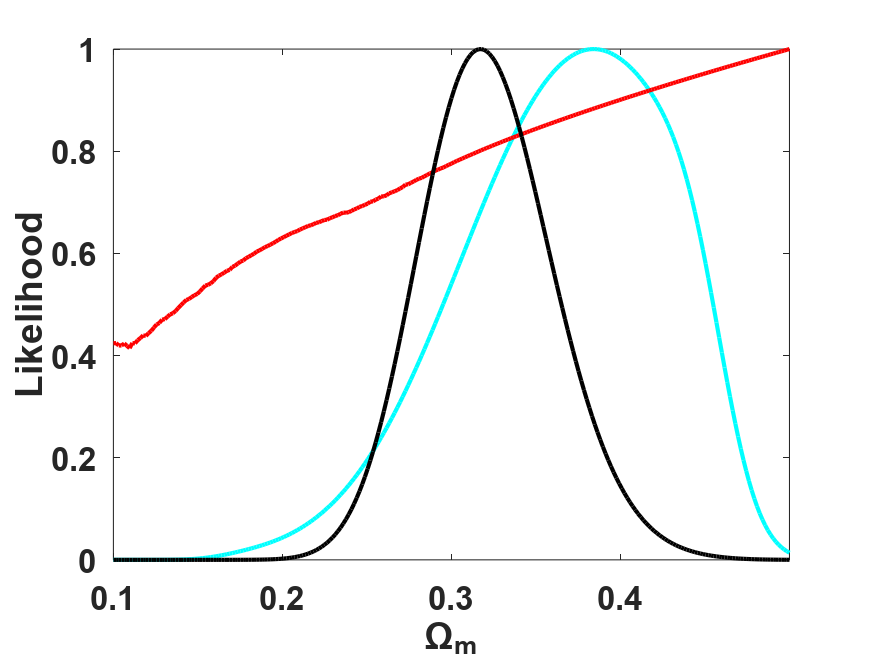}
\end{center}
\caption{\label{fig4}Constraints on the fractional cosmology model with $\Omega_\Lambda\neq0$. Blue and red curves show constraints from cosmology and temperature data, while black ones depict the combined constraints. The panels show the posterior marginalized likelihoods for $\alpha$ and the matter density.}
\end{figure*}

\section{Conclusions}\label{concl}

Our analysis shows that recent progress in measuring the temperature of the cosmic microwave background at various non-zero redshifts hasled to a data set which plays a significant role constraining fundamental physics and cosmology paradigms. In particular, we have shown that these measurements have a constraining power that is comparable to that of other background cosmology observables probing similar redshifts, specifically Type Ia supernovae and Hubble parameter measurements.

In next decade, the high-resolution ELT spectrograph \cite{HIRES}, ANDES, will significantly improve the sensitivity of these measurements in the deep matter era, while next-generation ground and possibly space CMB experiments \cite{CORE} will lead to comparable progress in lower-redshift Sunyaev-Zel'dovich effect measurements.Together, these will ensure that temperature measurements will remain a competitive probe of new physics.

More broadly, the fact that the observed temperature-redshift relation is consistent with the canonical prediction is further evidence for the fact that $\Lambda$CDM is a robust paradigm. While it is undoubtedly a phenomenological approximation to a yet unknown more fundamental model, any plausible alternative model must closely reproduce its low-redshift behaviour.

\section*{Acknowledgements}

This work was financed by FEDER---Fundo Europeu de Desenvolvimento Regional funds through the COMPETE 2020---Operational Programme for Competitiveness and Internationalisation (POCI), and by Portuguese funds through FCT - Funda\c c\~ao para a Ci\^encia e a Tecnologia in the framework of the project POCI-01-0145-FEDER-028987 and PTDC/FIS-AST/28987/2017. CJM also acknowledges FCT and POCH/FSE (EC) support through Investigador FCT Contract 2021.01214.CEECIND/CP1658/CT0001. The project that led to this work was started during AstroCamp 2020. AMV was partially supported by Ci\^encia Viva OCJF funds.

\bibliographystyle{model1-num-names}
\bibliography{alpha}
\end{document}